# Tunability of martensitic behavior through coherent nanoprecipitates and other nanostructures

Samuel Temple Reeve[a], Karthik Guda Vishnu[a], Alexis Belessiotis-Richards[b], and Alejandro Strachan[a][*]

[a]School of Materials Engineering and Birck Nanotechnology Center,

Purdue University, West Lafayette, Indiana 47906 USA

[b]Department of Materials,

Imperial College London, Exhibition Road, London SW7 2AZ, United Kingdom

## Abstract

Molecular dynamics simulations show that coherent precipitates can significantly affect the properties of martensitic transformations in $Ni_{63}Al_{37}$ alloys. The precipitates, consisting of non-martensitic $Ni_{50}Al_{50}$, modify the free energy landscape that governs the phase transformation and result in a significant reduction of the thermal hysteresis, at comparably minor expense of transformation strain, and modification of transformation temperatures. Importantly, this paper shows that free energy landscape engineering is possible with nanostructures potentially accessible through standard metallurgical processing routes. The atomistic-level nucleation and transformation mechanisms within the nanoprecipitate systems are explored and compared with epitaxial nanolaminates and nanowires. The simulations reveal three distinct regimes of transformation mechanisms and martensitic nanostructure as a function of volume fraction of the non-martensitic phase. Free energy landscape engineering is generally applicable and could contribute to the design of new shape memory alloys with novel properties, such as light weight alloys that operate at room temperature.

---

[*] Corresponding Author: strachan@purdue.edu

# 1. Introduction

Materials that exhibit shape memory and superelasticity are used in a wide variety of applications, including solid state actuators, biomedical devices, vibration damping, and in solid-state refrigeration [1–4]. These two phenomena originate in a type of solid to solid phase transformation referred to as martensitic, where a material transforms from a high-temperature, high-symmetry phase (austenite) to a lower symmetry phase (martensite) upon cooling or mechanical deformation. Importantly, these transformations occur without atomic diffusion. Materials design efforts, including compositional and microstructural optimization [5–9], have resulted in significant ability to tune transition temperatures, reduce thermal and mechanical hysteresis, and improve the reversibility of the transformation. In addition, mechanistic understanding of crystallographic relationships between the martensite and austenite phases has proved instrumental in understanding and engineering the phase transformations and derived properties [10,11]. Despite this progress, the need for additional development and new avenues to optimize the properties of this class of materials is exemplified by a new series of shape memory alloys (SMAs). These rare-earth Mg alloys have a significantly lower density than other SMAs, making them attractive for many applications. Unfortunately, their low transformation temperatures render them impractical [12]. In this paper, we explore how coherent precipitates in a martensitic matrix can be used to tune the transformation temperature and hysteresis using molecular dynamics (MD) simulations.

While there has been substantial research on the effect of precipitates in SMAs [1,13–15], recent experimental and theoretical work has shown that coherent integration of an appropriately chosen second phase into martensitic materials can open new dimensions to improve performance or achieve properties not otherwise accessible [16–19]. Notably, a recent study found the presence of coherent $Ti_2Cu$ nanoprecipitates in a TiNiCu SMA matrix led to significant improvements in the fatigue life and strength [17]. Theory and simulation have shown that coherent integration of a second phase can be used to modify the free energy landscape that governs the martensitic transformation and, thus, engineer desired properties. Indeed, significant ability to tune transformation temperature and reduce hysteresis using free energy landscape engineering (FELE) in coherent nanolaminates has been demonstrated via MD simulations [19]. MD simulations on $NiTi/Ni_3Ti_4$ systems showed that precipitates can reduce transformation stress, at the expense of transformation strain [20], while the opposite arrangement of phases (NiTi precipitates in a non-transforming matrix) resulted in modest modification of thermal and mechanical transformations [21]. In addition, ultra-low stiffness metals have been demonstrated using high-fidelity MD simulations of coherent core/shell nanowires and nanolaminates by choosing a second phase capable of stabilizing the martensitic phase in a thermodynamically unstable state of negative stiffness [18]. The concept of FELE is exemplified graphically in Figure 1, where a coherent, linear-elastic second phase (blue line) modifies the martensitic free energy landscape (red line), making lower hysteresis possible by reducing the transformation energy barrier, as well as stabilizing the austenite phase with respect to the martensite. To date, the ability to tune the transformation temperature, hysteresis, and achieve ultra-low stiffness has only been demonstrated in nanolaminate and core/shell nanowire geometries [18,19], both of which are difficult to achieve experimentally. In addition, open questions remain about the operating mechanisms. Inspired by the growing number of aforementioned martensitic materials incorporating useful coherent precipitates, this paper explores FELE via coherent nanoprecipitates to tune the temperature and hysteresis associated with the martensitic transformation in $Ni_xAl_{1-x}$ alloys. We compare results for coherent precipitates with nanolaminates and core/shell nanowire composite systems, showing tunable properties for all geometries and how the volume fraction of the precipitates affects the structural transformation and the resulting martensitic microstructures. We establish correlations between the underlying free energy landscape and the thermal behavior of the material; this is dominated by either the reduction of the barrier associated with the transformation or stabilization of austenite.

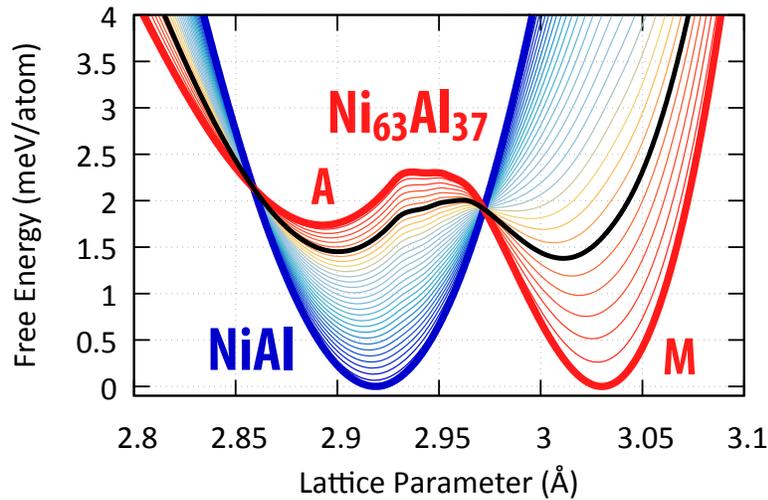

Figure 1: Free energy landscape engineering example combining martensitic ($Ni_{63}Al_{37}$, red) and non-martensitic (NiAl, blue) components. Both are bulk systems at 750 K, strained biaxially, described in Section 2, interpolated to show possible effects of coherent integration.

The paper is organized as follows. Simulation details are given in Section 2: the interatomic model, system creation, thermal cycling, structural characterization, and free energy calculation. Section 3.1 shows the thermal cycling results for all geometries and the martensite nucleation for the precipitate systems. Next, trends in thermal hysteresis reduction and corresponding transformation strain loss are discussed in Section 3.2. Opposing trends for transformation temperatures are finally explained with the free energy landscapes in Section 4, followed by conclusions and continuing work in Section 5.

## 2. Simulation details

### 2.1 Model systems and interatomic potentials

Quenched $Ni_xAl_{1-x}$ alloys display a martensitic transformation and shape memory between 60 and 65 at. % Ni [22]. Here, 63 at. % Ni – 37 at. % Al systems were created starting from perfect B2 crystals with Al sites randomly swapped for Ni in the desired region. In all cases, the non-martensitic second phase is perfect B2 $Ni_{50}Al_{50}$. The interatomic interactions were described with a potential developed by Farkas et al. [23], parametrized based on experimental properties of FCC Ni and Al, $L1_2$ $Ni_3Al$, B2 NiAl, *Cmmm* $Ni_5Al_3$, and $L1_0$ martensite. The potential has been shown to accurately capture a cubic austenite to monoclinic martensite transformation, where two lattice parameters expand and one contracts, with reasonable transformation temperatures for the correct range of Ni compositions. The model and motivation of the approach are described in more detail in prior publications [19,18,24,25]. The Farkas model has been used extensively to investigate many features of martensitic transformations from the atomic scale including simulation size and sample variability [24], surface [26,27], and grain size [25] effects. All simulations were performed using LAMMPS [28].

### 2.2 Structure generation and thermal cycling

We studied three different configurations: a spherical precipitate in a bulk matrix, nanolaminates, and both core and shell nanowires. All bulk systems (nanoprecipitate and nanolaminate) were created with 1,024,000 atoms, replicating the B2 unit cell 80 times in each direction, resulting in initial linear dimensions between 23.2 nm and 23.3 nm. The nanowires had initial lengths between 34.0 nm and 34.6 nm (120 unit cells) and radii between 5.88 nm and 5.92 nm (20 unit cells) with 302,520 atoms; these systems are periodic only along their axes. Systems with NiAl precipitates in a $Ni_{63}Al_{37}$ matrix will be denoted *driving matrix* (D-Matrix), referring to the fact that the off-stoichiometric phase prefers to transform and can force the second phase to transform as well. D-Matrix systems are built with a central sphere of perfect B2 NiAl of the desired radius. Systems with the phases reversed are referred to as *driving* *precipitate* (D-Precipitate). Similarly, nanolaminates with layered regions, as well as D-Core and D-Shell nanowires with cylindrical regions were built. For all geometries, numerous volume fractions of NiAl and $Ni_{63}Al_{37}$ were simulated, beginning with bulk $Ni_{63}Al_{37}$ until the transformation was completely suppressed. For the nanoprecipitates, volume fractions greater than 52% of the system are not geometrically possible. Systems are referred to with their volume fraction of the NiAl phase.

Each system was equilibrated at 800 K for 100 ps, cooled to 25 K at a rate of $1 \cdot 10^{11}$ K/s, and heated above the austenite finish ($A_f$) temperature at the same rate. Throughout, an NPT ensemble was used with a thermostat coupling constant of 0.1 and barostat coupling constant of 0.1. During equilibration and thermal cycling all directions and angles were left free.

### 2.3 Structural Analysis

All visualization and structural analysis was performed using OVITO [29]. The polyhedral template matching (PTM) algorithm [30] implemented in OVITO was used for structural identification of austenite and martensite. For each atom, the local neighbor structure is compared to templates for common crystals (FCC, BCC, HCP, etc.) and the root mean square (RMS) error calculated. Each atom is then identified as the structure with the lowest RMS error. For this work, atoms identified as BCC are referred to as austenite, HCP as martensite, and FCC as stacking faults. If no mapping exists to any of the chosen templates, the atom is listed as "other". In addition, an RMS error cutoff of 0.12 was chosen, above which atoms were also considered "other". As a method based on topology, rather than bond cutoffs (e.g. common neighbor analysis), it is more accurate for the high strain and high temperature systems investigated here.

### 2.4 Free energy landscape calculations

After determining the martensite start/finish ($M_s$/$M_f$) and austenite start/finish ($A_s$/$A_f$) transformation temperatures from the cooling-heating cycle, free energy landscapes were calculated at $T_0 = (A_f + M_f)/2$ for each system. Systems without a discernable $M_f$ were considered to have a $T_0$ of 25K. Starting from the stable, unstrained, transformed state (from the heating curve), the structures were equilibrated for 50 ps and then strained in tension and compression. Because it is an isothermal process, the entropic term goes to zero and it is possible to integrate the free energy from the stress and strain as:

$$dF = V(\sigma_{xx}d\epsilon_{xx} + \sigma_{yy}d\epsilon_{yy} + \sigma_{zz}d\epsilon_{zz})$$

This constitutes a "drag method" for free energy calculation, where the system is forced to follow a specific path. This differs in comparison to other methods for free energy calculation for solid state transformations discussed in Ref. [31], notably the generalized solid state nudged elastic band. Nanoprecipitate and nanolaminate landscapes were calculated with biaxial strain controlling the two longer lattice directions at a rate of $1 \cdot 10^9$ s$^{-1}$, while nanowire free energy landscapes were performed uniaxially along the wire axis at a rate of $1 \cdot 10^8$ s$^{-1}$. The slower rate for the wires was necessary due to the smaller system size and higher noise with the free surfaces. For biaxial strain, most systems showed pseudo-tetragonal behavior such that equal strain for the two longer directions goes through the cubic state. In systems that did not, the two controlled directions were strained directly to the cubic state (from the cooling curve at that temperature) and then strained equally. An NPT ensemble was again used, with only unstrained, periodic directions and angles relaxed. For each case, the unstrained directions contribute negligibly to the free energy.

## 3. Role of the second phase on martensitic transformation

### 3.1 Thermal transformation and structural regimes

The previous study on nanolaminate systems [19] showed significant reduction of thermal hysteresis through FELE. New simulations of the thermally induced martensitic transformation upon cooling and austenitic transition upon reheating for laminates are shown in Figure 2a as a reference. As each system is cooled from 800K, the martensitic transformation is marked by rather abrupt lattice parameter changes with temperature which deviate significantly from thermal contraction; two lattice parameters expand and one contracts. Similarly, when heated from 25K, the eventual transition back to austenite occurs when the lattice parameters return to nearly equal values (the cubic phase). This temperature loop is highlighted with arrows for the bulk alloy in Fig. 2a. As previously reported, we observe a significant reduction in transformation hysteresis with increasing volume fraction of the NiAl component. As could be expected from Figure 1, this stems from the engineered free energy landscape with a reduction in the energy barrier separating these two phases. Results for cooling and heating of the nanoprecipitate systems show that the martensitic matrix systems exhibit similar trends, Figure 2b. This is an important result as it shows that FELE is possible in geometries accessible via standard metallurgical processing routes. We find a gradual reduction in hysteresis accompanied by a minimal reduction in transformation strain as we increase the volume fraction of the NiAl precipitate up to 30%. Increasing the precipitate volume to 40% leads to a sharp reduction in hysteresis and transformation strain; as discussed below, this indicates a change in transformation mechanism. Interestingly, the D-Precipitate systems did not transform thermally at any volume fraction studied, up to 40 at. % Ni$_{63}$Al$_{37.}$ For the precipitate configurations we are studying, spheres in a simple cubic arrangement, a maximum volume fraction of ~52% can be achieved before the precipitates come in contact with its periodic images. We note that other precipitate arrangements and shapes that enable a higher volume fraction of the transforming phase may result in transformation. For completeness, Figure 2 also includes results for nanowires in both the D-Core and D-Shell configurations. This configuration produces similar trends, but with significantly more modification of the $M_s$ temperatures (see Figures 2c-d). In all cases the reduction in thermal hysteresis is accompanied by a reduction in transformation strain.

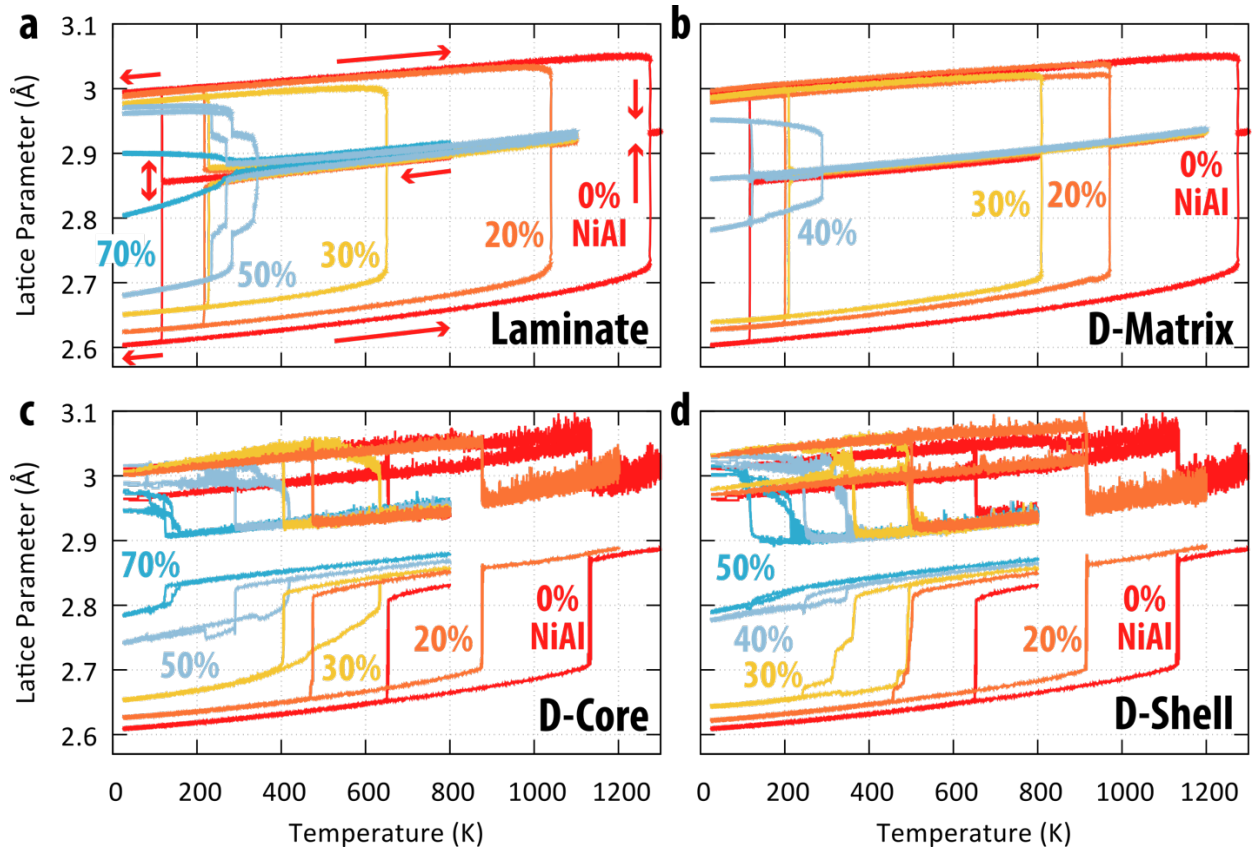

**Figure 2: Lattice parameter evolution during cooling-heating cycle in a) nanolaminates, b) D-Matrix nanoprecipitates, c) D-Core nanowires, and d) D-Shell nanowires. Temperature path shown with arrows in a).**

Structural analysis of the MD trajectories provide insight into the temperature dependence of the lattice parameters for the various cases. We first explore the martensitic transformation of two D-Matrix nanoprecipitate systems (40 and 30 at. % NiAl) upon cooling to highlight the effect of the second phase on the martensitic transformation, see Figure 3. Atoms are colored according to their local environments using the PTM method, classified into B2 austenite (blue), monoclinic martensite (red), stacking faults (green), and other defects (white), with the front half of atoms in the simulation cell removed for clarity. NiAl and $Ni_{63}Al_{37}$ phases are also distinguished as light and dark shades of these colors, respectively. For both volume fractions, the systems start as fully austenite. The left panels in Figure 3 show the nucleation of the martensitic phase, immediately below the $M_s$. Both systems behave similarly, exhibiting multiple martensitic domains with different orientations that nucleate entirely within the $Ni_{63}Al_{37}$ region near the surface of the precipitate. The second view of the snapshots in Figs. 3a,c shows periodic replicas to highlight the various domains (see arrows); note that the orientation of the stacking faults (in green) can be used to assess the orientation of the martensite variant. This preferential nucleation contrasts the behavior of the bulk $Ni_{63}Al_{37}$ system where domains nucleate throughout the system. For the 30 at. % D-Matrix, as well as cases with lower volume fractions of NiAl, the NiAl nanoprecipitate eventually transforms into the martensitic phase, Fig. 3b, and a single orientation overtakes the entire sample. This process underlies the initial regime of gradual reduction in hysteresis and minimal reduction in transformation strain in Fig. 2b. While the nucleation of the martensite phase is similar for the 30% and 40% samples, only the matrix transforms for the 40 at. % D-Matrix (Fig. 3d) with the precipitate remaining in the austenite phase. The precipitate then limits the interaction between the multiple initial domains and we observe multiple persistent domains instead of a single domain overtaking the entire sample as described above. This partial transformation and multi-domain structure explain the drastically reduced hysteresis and strain in Fig. 2 for the 40 at. % system (including one lattice parameter that does not significantly change).

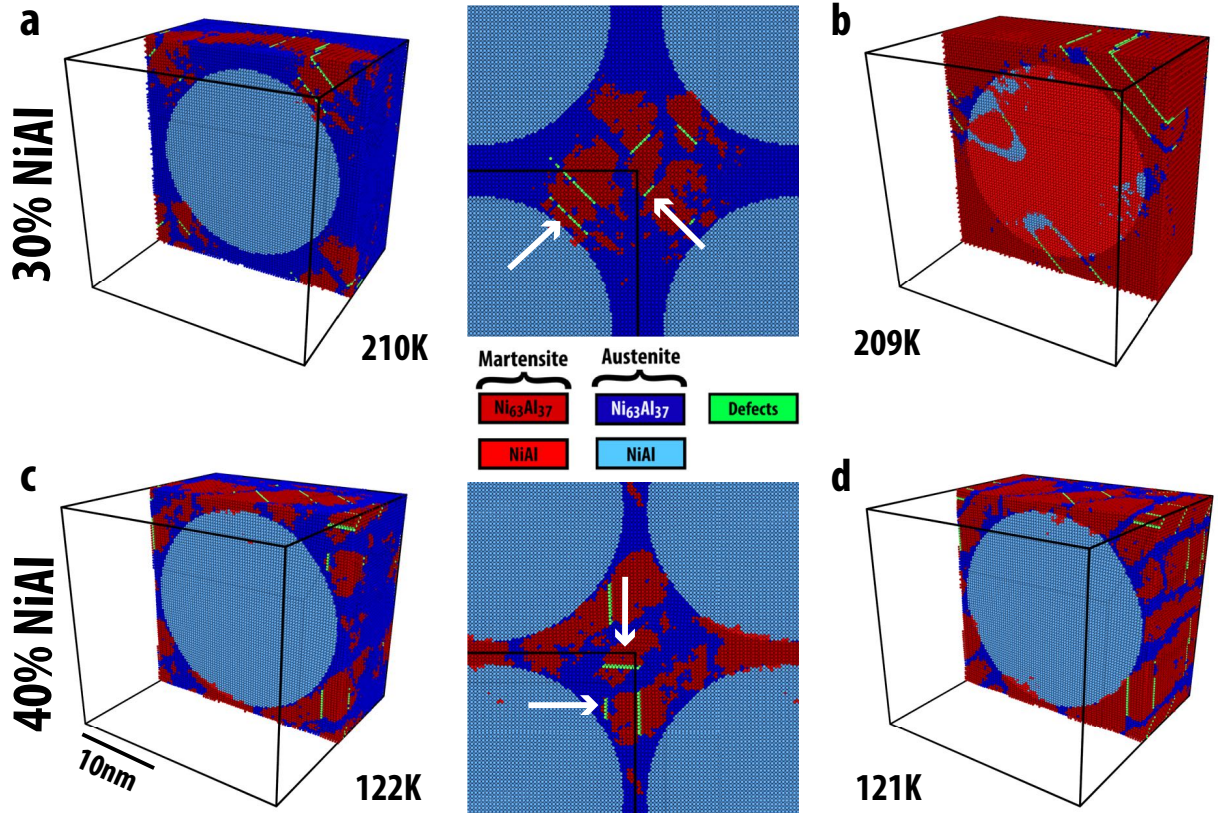

Figure 3: Martensitic nucleation in D-Matrix nanoprecipitates just below $M_s$ with 30 at. % NiAl at a) 210 K and b) 209 K ($M_s$ 211K) and 40 at. % NiAl at c) 122K and d) 121K ($M_s$ 128K). Front half of atoms are removed for clarity in all; atomic coloring is detailed in the center inset. Right snapshot in a) and c) show periodic replications of the systems to highlight two martensitic domains (with arrows) in each.

Turning our attention to the overall processes that govern the transformation in the various configurations studied, Figure 4 shows atomistic snapshots after cooling to 25 K for each geometry and several volume fractions. The front half of the D-Matrix, D-Core, and D-Shell systems are removed for clarity in Fig. 4. To complement the analysis, Figure 5 shows the evolution of the fraction of atoms classified as belonging to the martensite phase, calculated using PTM, during the cooling and heating process. We find three distinct regimes of transformation with respect to the volume fraction of the non-transforming phase, present in all geometries; the structures shown in Fig. 4 correspond to the highest volume fraction NiAl in each regime. The first regime occurs for the lowest volume fractions of the non-transforming NiAl phase and is characterized by the entire system transforming to martensite. This process results in minimal reduction in transformation strain as the $Ni_{63}Al_{37}$ phase forces the NiAl phase to transform. This is shown both in the top row of Fig. 4 with complete transformation to martensite (excluding domain boundaries in some systems), as well as the abrupt, almost vertical transformations in Fig. 5. The second regime involves intermediate fractions of the NiAl phase; we observe complete transformation of the $Ni_{63}Al_{37}$ region, but only partial transformation of the NiAl (middle row of Fig. 4). This results in a two-step transformation process (see, for example, Fig. 5a, 50 at.%). As expected, transformation occurs first in the $Ni_{63}Al_{37}$ phase and is subsequently followed by transformation of the non-martensitic NiAl. The NiAl region in these systems transforms less abruptly and incompletely; thus, there is a significant decrease in transformation strain. Finally, the third regime takes place at large volume fractions of NiAl where transformation is completely suppressed in the non-martensitic NiAl (bottom of Fig. 4). These systems also exhibit a gradual transformation of the $Ni_{63}Al_{37}$ region, as with the NiAl region in the previous regime. Due to the presence of free surfaces in the nanowire, the geometric constraints of the martensite and austenite coexistence are less significant. The nanowires therefore exhibit a wider range of transformation temperatures, with less clear distinction between regimes. The martensitic transformation is entirely suppressed with the highest NiAl volume fractions, above 80, 70, and 50 at. % for laminate, D-Core, and D-Shell systems, respectively; however, the D-Matrix nanoprecipitate systems cannot geometrically be created at high enough NiAl volume fractions to show behavior in the partially suppressed third regime.

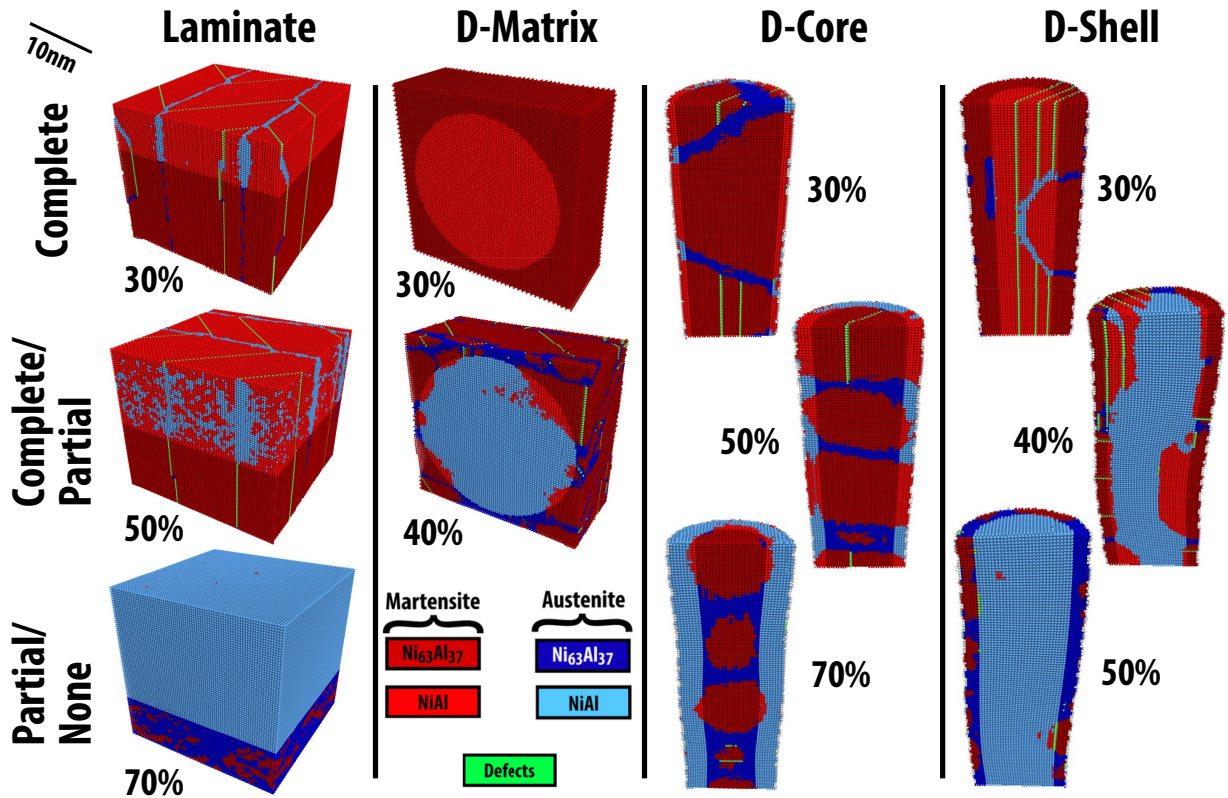

**Figure 4:** Atomic structure at 25K for structures representative of each regime in a) nanolaminates, b) D-Matrix nanoprecipitates, c) D-Core nanowires, and d) D-Shell nanowires. Atomic coloring is described at bottom.

The evolution of the martensitic fraction in the systems during cooling and heating shown in Fig. 5 correlate with the lattice parameter changes shown in Fig. 2. Note that the smooth reduction in martensite fraction at high temperature is partially due to thermal noise which leads to an increase in unidentifiable atomic coordination ("other" atoms). Increasing the NiAl volume fraction in all geometries brings the two characteristic temperatures ($A_f$ and $M_s$) closer together, demonstrating reduced hysteresis. Shifts in the transformation temperature are also captured, particularly in volume fractions at and above 40 at. % NiAl.

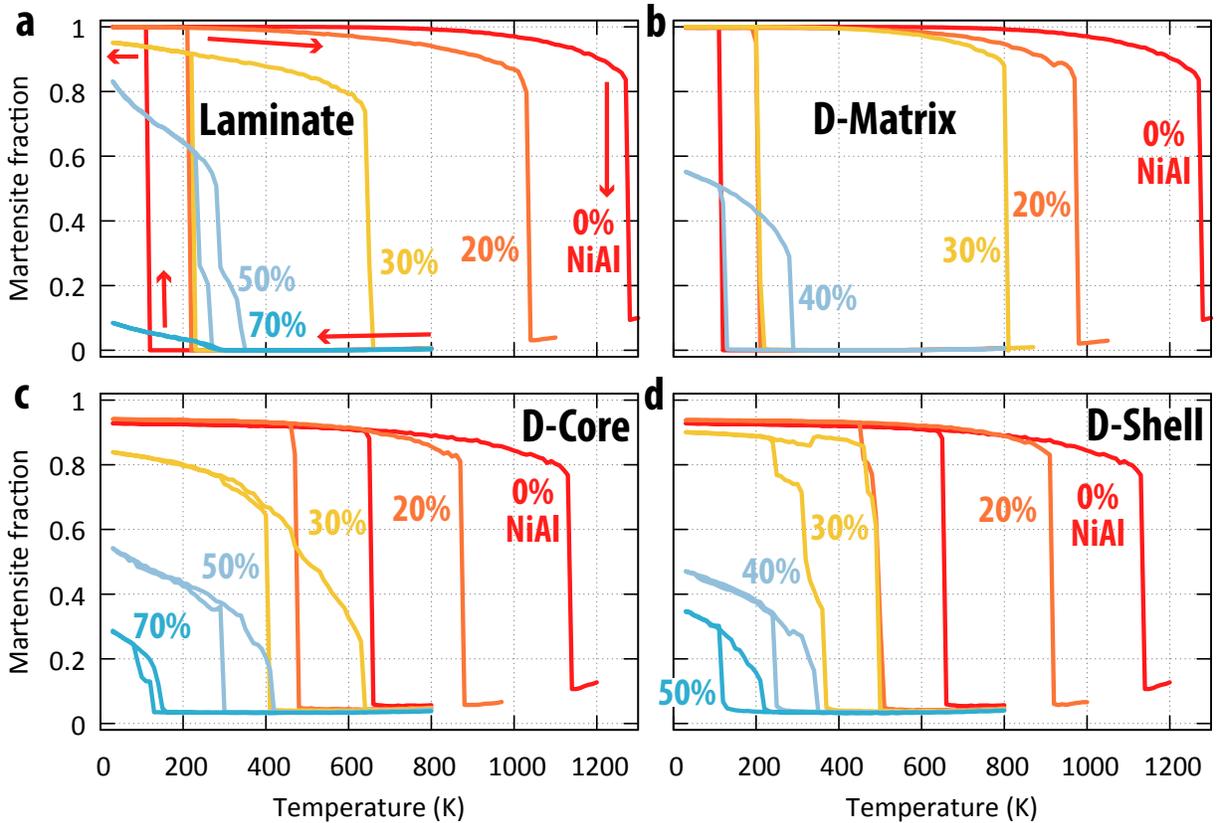

**Figure 5: Martensite evolution (including stacking faults) during the cooling-heating cycle in a) nanolaminates, b) D-Matrix nanoprecipitates, c) D-Core nanowires, and d) D-Shell nanowires. Temperature path shown with arrows in a).**

3.2 Tunability of hysteresis and martensitic transformation temperatures

For many applications of SMAs, minimization of thermal hysteresis and maximization of transformation strain is desirable. We therefore compare the effect of adding a NiAl second phase on the purposeful reduction in hysteresis and unavoidable reduction in strain for all geometries in Figure 6. Hysteresis is defined as ($A_f$ – $M_s$) and transformation strain as the change in lattice parameter from $M_s$ to 25K in each direction. The ideal behavior would lie in the lower right corner of the plots and we find that all geometries display the preferred balance of a more substantial hysteresis reduction than the accompanying transformation strain reduction. The uncertainty in the results can be estimated by the 95% confidence interval, shown with error bars in both hysteresis and strain, from three independent simulations of the bulk system (top right, Fig. 6). In addition, the uncertainty can be inferred from the scatter in similar volume fractions, where intermediate volume fractions show largest variability as in Ref. [18]. The volumetric strain in the wire systems was obtained from the change in length (obtained from the periodic simulation cell length) and the transverse strain obtained by sectioning the wire into ten slabs and averaging. We note that volumetric strain in the wire systems is of little practical importance. The lowest volume fraction NiAl nanolaminate and nanoprecipitate systems show strong reduction of hysteresis with minimal loss of strain (top right, Fig.6). For the highest NiAl volume fractions, all geometries produce small, but finite strain with essentially zero thermal hysteresis; this could be potentially useful for niche low strain, ultra-low hysteresis applications.

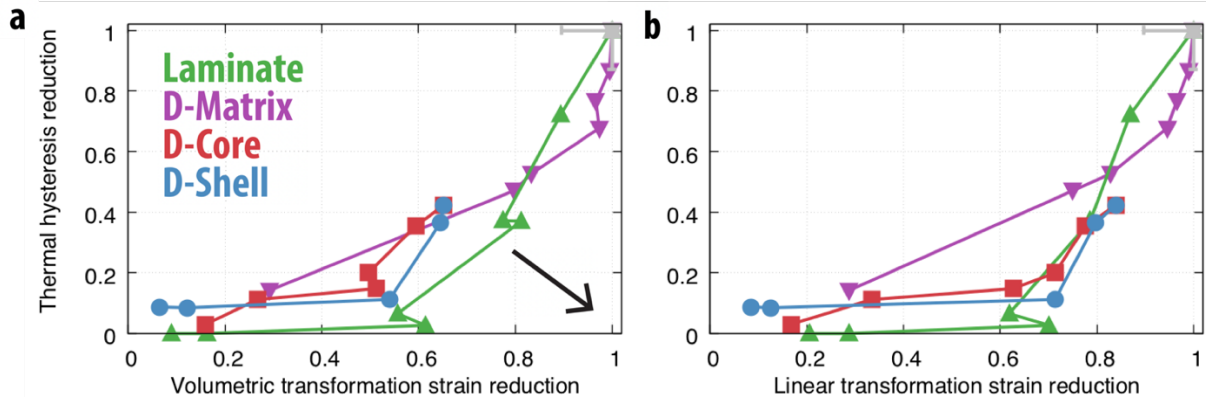

**Figure 6: Comparison of a) volumetric and b) linear transformation strain reduction vs. thermal hysteresis reduction, all scaled by the properties of the bulk $Ni_{63}Al_{37}$ alloy (shown with 95% confidence intervals from three independent samples in grey). Points are ordered and connected for each geometry by NiAl volume fraction.**

## 4. Free energy landscape engineering

Underlying the effect of the NiAl phase on thermal hysteresis presented above is the effect on the individual transformation temperatures, $M_s$ and $A_f$. Figure 7 shows the dependence of the transformation temperatures on NiAl content. As expected, the $A_f$ temperature decreases with increasing volume fraction of NiAl for all geometries, see Fig. 7a. Recalling the free energy landscape changes from Fig. 1, this is consistent with both the stabilization of the austenite phase over a wider temperature range and the reduction in the free energy barrier that makes it easier to transform from martensite back to austenite. However, the effect of the second phase on the $M_s$ temperature (Fig. 7b) shows intriguing, opposite, trends in nanowire geometries as compared with nanolaminate and nanoprecipitates. While the addition of NiAl would be expected to similarly reduce the austenite to martensite transformation barrier, leading to an increase in Ms, the stabilization of the austenite phase would have the opposite effect, i.e. to depress the Ms temperature. This competition between mechanisms is therefore able to explain both trends. The nanolaminate and D-Matrix nanoprecipitates show increase in $M_s$ with increasing non-martensitic NiAl volume fraction; this indicates that the reduction in energy barrier dominates. The decrease in $M_s$ for the nanowires conversely indicates that stabilization of the austenite phase prevails.

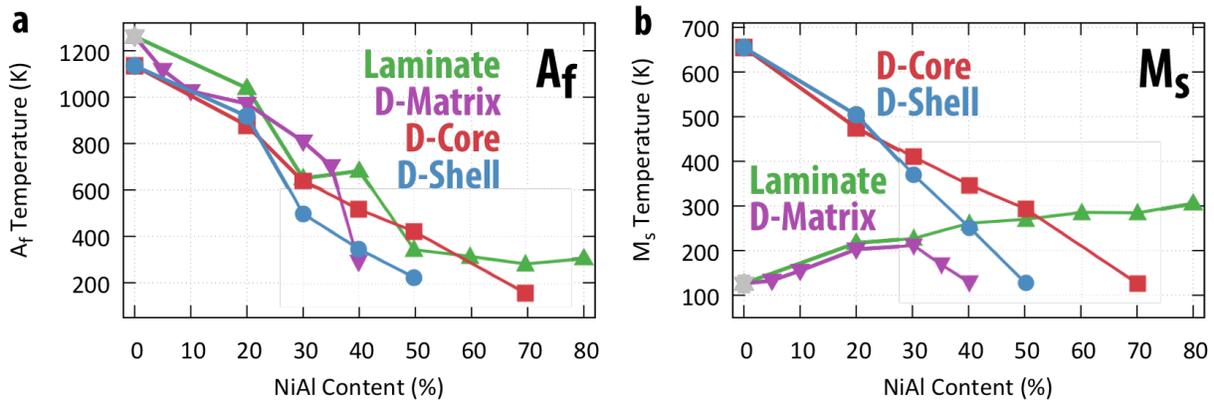

**Figure 7: Transformation temperatures for all systems with a) austenite finish ($A_f$) matching trends across geometries and b) martensite start ($M_s$) contrasting geometries. Confidence intervals of 95% for three independent bulk samples shown in grey.**

In order to confirm our understanding of the trends in hysteresis and transformation temperature, we computed the free energy landscapes for all geometries and volume fractions at relevant temperatures. As described in Section 2, these landscapes relate the free energy to the strain used to drive the transformation; we use biaxial strain for the nanolaminates and nanoprecipitates and uniaxial deformation for the nanowires, see Figure 8. It should be noted that in the nanowire cases, driven by surface energies, the preferred martensite variant aligns its short direction along the axis of the wire. Further, the lattice parameter along this axis is the only possible progress variable for computing the free energy landscape in these systems and is therefore used in Figs. 8c-d. For these systems, the martensite therefore lies at lower values of lattice parameter (i.e. to the left of the austenite).

For both the nanolaminates and D-Matrix nanoprecipitates, the energy barrier associated with the transformation from martensite to austenite, going from right to left in Figures 8a-b, is reduced by the incorporation of the coherent NiAl phase, as designed. This explains the reduction in hysteresis associated with the transformation and the stabilization of the austenite phase at the expense of the martensite, resulting in the reduction in $A_f$ (Figure 7a). The increase in $M_s$ with increasing fraction of NiAl for these systems also matches the observed reduction in barrier associated with the austenite to martensite transformation (left to right), which for these systems more than compensates the increased stability of the austenite (Fig. 8a).

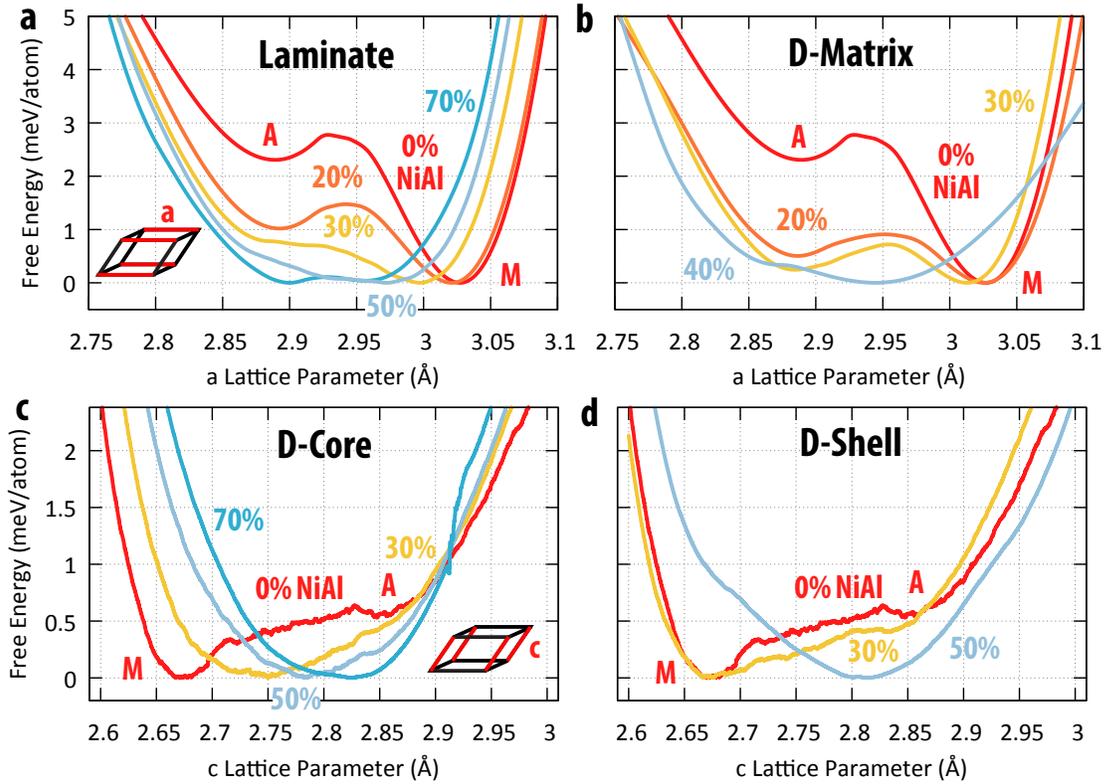

**Figure 8: Free energy landscapes for various NiAl fraction systems at $T_0 = (A_f + M_f)/2$ in a) nanolaminates, b) D-Matrix nanoprecipitates, c) D-Core nanowires, and d) D-Shell nanowires.**

For the nanowire cases, the barrier also reduces significantly from martensite to austenite (left to right) with increasing volume fraction NiAl, matching the shared trend of reduced $A_f$. However, primarily due to the free surface, all nanowires exhibit slightly increased or unchanged barriers for the reverse path, from austenite to martensite (right to left), with increasing fraction of NiAl (Fig. 8c-d), where even the bulk wire shows a very small barrier. Thus, austenite stabilization is dominant in the wires, yielding the opposite $M_s$ trend with volume fraction. Supplementary Figure 1 shows that the austenite is stable in the nanowires (with zero curvature at those states), more easily accessible beginning from that state.

In agreement with the systems displaying lowest thermal hysteresis, the free energy landscapes for the 40-70 at. % NiAl exhibit distinct features. For these volume fractions, the landscapes are optimally balanced such that the free energy profile is nearly flat for both the stable and metastable states. FELE results in stabilization of new states of coexistence containing both austenite and martensite, states in between the individual martensite and austenite phases in Fig. 1. Further, the flattening of the energy landscapes agree with the significant changes in elastic behavior from landscape engineering [18].

In contrast to the engineered free energy landscapes here, bulk landscapes for the $Ni_{63}Al_{37}$ phase at various temperatures are shown in Supplementary Figure 2. For future work, free energy landscapes interpolated from the bulk phases could be used as a first step in the identification of appropriate phases for specific properties and landscapes of the specific structures of interest then used to narrow down options even further, significantly reducing the computational cost to predicting useful combinations of phases.

## 5. Conclusions

We use MD simulations to show that the incorporation of coherent precipitates in a martensitic matrix modifies the underlying free energy landscape that governs the phase transformation in the alloy. In the cases studied, the second phase resulted in the stabilization of the austenite phase and reduction of the transformation energy barrier. Thus, the presence of the precipitate resulted in a reduction of the hysteresis associated with the martensitic transition and modified the transformation temperatures. These property changes can be understood in terms of the competition between the stabilization of the austenite, resulting in a reduction of both martensitic and austenitic transformation temperatures, and the reduction in energy barrier associated with the transformation that reduced hysteresis, increasing the martensitic temperature and decreasing the austenitic transition temperature. The reduction in hysteresis is desirable for applications involving actuation but it comes accompanied by an undesirable reduction in transformation strain. We find that the relative reduction in hysteresis surpasses that in strain. Detailed analysis of the MD trajectories identified three distinct regimes in terms of how the presence of the NiAl phase affects the martensitic transformation and microstructure. The first regime shows comparable behavior to the bulk alloy, with complete transformation as the martensitic phase forces the NiAl to transform. Within the second regime the NiAl phase only partially transforms, with many systems retaining multiple domains, with near equal balance of the two phases. The martensitic region transforms only partially for the third regime, again in multi-domain structures, and with little to no transformation of the NiAl.

The principles of FELE can be generally applied to modify the properties of a martensitic material as long as an appropriate, coherent, second phase can be identified. FELE could be utilized to identify ideal second phases in shape memory alloys in conjunction with materials informatics and thermodynamic approaches for composition and processing condition optimization [5–8,32,33]. One potential application of FELE could be to increase the operating temperature of new light weight SMAs [12] by stabilizing the martensite phase.

# Acknowledgements

This work was supported by the United States Department of Energy Basic Energy Sciences (DoE-BES) program under Program No. DE-FG02-07ER46399 (Program Manager John Vetrano). Computational resources from nanoHUB and Purdue University are gratefully acknowledged.

# References

[1]     K. Otsuka, X. Ren, Physical metallurgy of Ti–Ni-based shape memory alloys, Prog. Mater. Sci. 50 (2005) 511–678. doi:10.1016/j.pmatsci.2004.10.001.

[2]     N.. Morgan, Medical shape memory alloy applications—the market and its products, Mater. Sci. Eng. A. 378 (2004) 16–23. doi:10.1016/j.msea.2003.10.326.

[3]     J. Van Humbeeck, Non-medical applications of shape memory alloys, Mater. Sci. Eng. A. 273 (1999) 134–148.

[4]     K. Otsuka, X. Ren, Martensitic transformations in nonferrous shape memory alloys, Mater. Sci. Eng. A. 273 (1999) 89–105.

[5]     D. Xue, D. Xue, R. Yuan, Y. Zhou, P.V. Balachandran, X. Ding, J. Sun, T. Lookman, An informatics approach to transformation temperatures of NiTi-based shape memory alloys, Acta Mater. 125 (2017) 532–541. doi:10.1016/j.actamat.2016.12.009.

[6]     D. Xue, P.V. Balachandran, J. Hogden, J. Theiler, D. Xue, T. Lookman, Accelerated search for materials with targeted properties by adaptive design, Nat. Commun. 7 (2016) 11241. doi:10.1038/ncomms11241.

[7]     R. Zarnetta, R. Takahashi, M.L. Young, A. Savan, Y. Furuya, S. Thienhaus, B. Maaß, M. Rahim, J. Frenzel, H. Brunken, Y.S. Chu, V. Srivastava, R.D. James, I. Takeuchi, G. Eggeler, A. Ludwig, Identification

of Quaternary Shape Memory Alloys with Near-Zero Thermal Hysteresis and Unprecedented Functional Stability, Adv. Funct. Mater. 20 (2010) 1917–1923. doi:10.1002/adfm.200902336.

[8] J. Cui, Y.S. Chu, O.O. Famodu, Y. Furuya, J. Hattrick-Simpers, R.D. James, A. Ludwig, S. Thienhaus, M. Wuttig, Z. Zhang, I. Takeuchi, Combinatorial search of thermoelastic shape-memory alloys with extremely small hysteresis width, Nat. Mater. 5 (2006) 286. doi:10.1038/nmat1593.

[9] Y. Song, X. Chen, V. Dabade, T.W. Shield, R.D. James, Enhanced reversibility and unusual microstructure of a phase-transforming material, Nature. 502 (2013) 85. doi:10.1038/nature12532.

[10] Y. Gao, L. Casalena, M.L. Bowers, R.D. Noebe, M.J. Mills, Y. Wang, An origin of functional fatigue of shape memory alloys, Acta Mater. 126 (2017) 389–400. doi:10.1016/j.actamat.2017.01.001.

[11] K. Bhattacharya, S. Conti, G. Zanzotto, J. Zimmer, Crystal symmetry and the reversibility of martensitic transformations, Nature. 428 (2004) 55. doi:10.1038/nature02378.

[12] Y. Ogawa, D. Ando, Y. Sutou, J. Koike, A lightweight shape-memory magnesium alloy, Science. 353 (2016) 368–370. doi:10.1126/science.aaf6524.

[13] J. Khalil-Allafi, A. Dlouhy, G. Eggeler, Ni4Ti3-precipitation during aging of NiTi shape memory alloys and its influence on martensitic phase transformations, Acta Mater. 50 (2002) 4255–4274. doi:10.1016/S1359-6454(02)00257-4.

[14] S. Kajiwara, D. Liu, T. Kikuchi, N. Shinya, Remarkable improvement of shape memory effect in Fe-Mn-Si based shape memory alloys by producing NbC precipitates, Scr. Mater. 44 (2001) 2809–2814. doi:10.1016/S1359-6462(01)00978-2.

[15] H.E. Karaca, S.M. Saghaian, G. Ded, H. Tobe, B. Basaran, H.J. Maier, R.D. Noebe, Y.I. Chumlyakov, Effects of nanoprecipitation on the shape memory and material properties of an Ni-rich NiTiHf high temperature shape memory alloy, Acta Mater. 61 (2013) 7422–7431. doi:10.1016/j.actamat.2013.08.048.

[16] S. Jiang, H. Wang, Y. Wu, X. Liu, H. Chen, M. Yao, B. Gault, D. Ponge, D. Raabe, A. Hirata, M. Chen, Y. Wang, Z. Lu, Ultrastrong steel via minimal lattice misfit and high-density nanoprecipitation, Nature. 544 (2017) 460–464. doi:10.1038/nature22032.

[17] C. Chluba, W. Ge, R. Lima de Miranda, J. Strobel, L. Kienle, E. Quandt, M. Wuttig, Ultralow-fatigue shape memory alloy films, Science. 348 (2015) 1004–1007. doi:10.1126/science.1261164.

[18]    S.T. Reeve, A. Belessiotis-Richards, A. Strachan, Harnessing mechanical instabilities at the nanoscale to achieve ultra-low stiffness metals, Nat. Commun. 8 (2017). doi:10.1038/s41467-017-01260-6.

[19]    K. Guda Vishnu, A. Strachan, Shape memory metamaterials with tunable thermo-mechanical response via hetero-epitaxial integration: A molecular dynamics study, J. Appl. Phys. 113 (2013) 103503. doi:10.1063/1.4794819.

[20]    P. Chowdhury, L. Patriarca, G. Ren, H. Sehitoglu, Molecular dynamics modeling of NiTi superelasticity in presence of nanoprecipitates, Int. J. Plast. 81 (2016) 152–167. doi:10.1016/j.ijplas.2016.01.011.

[21]    S.B. Maisel, W.-S. Ko, J.-L. Zhang, B. Grabowski, J. Neugebauer, Thermomechanical response of NiTi shape-memory nanoprecipitates in TiV alloys, Phys. Rev. Mater. 1 (2017) 033610. doi:10.1103/PhysRevMaterials.1.033610.

[22]    P.S. Khadkikar, I.E. Locci, K. Vedula, G.M. Michal, Transformation to Ni5Al3 in a 63.0 At. Pct Ni-Al alloy, Metall. Trans. A. 24 (1993) 83–94. doi:10.1007/BF02669606.

[23]    D. Farkas, B. Mutasa, C. Vailhe, K. Ternes, Interatomic potentials for B2 NiAl and martensitic phases, Model. Simul. Mater. Sci. Eng. 3 (1995) 201. doi:10.1088/0965-0393/3/2/005.

[24]    K.R. Morrison, M. Cherukara, K. Guda Vishnu, A. Strachan, Role of atomic variability and mechanical constraints on the martensitic phase transformation of a model disordered shape memory alloy via molecular dynamics, Acta Mater. 69 (2014) 30–36. doi:10.1016/j.actamat.2014.02.001.

[25]    K.R. Morrison, M.J. Cherukara, H. Kim, A. Strachan, Role of grain size on the martensitic transformation and ultra-fast superelasticity in shape memory alloys, Acta Mater. 95 (2015) 37–43. doi:10.1016/j.actamat.2015.05.015.

[26]    Z. Zhang, X. Ding, J. Sun, T. Suzuki, T. Lookman, K. Otsuka, X. Ren, Nonhysteretic Superelasticity of Shape Memory Alloys at the Nanoscale, Phys. Rev. Lett. 111 (2013) 145701. doi:10.1103/PhysRevLett.111.145701.

[27]    H. Zong, Z. Ni, X. Ding, T. Lookman, J. Sun, Origin of low thermal hysteresis in shape memory alloy ultrathin films, Acta Mater. 103 (2016) 407–415. doi:10.1016/j.actamat.2015.10.033.

[28]    S. Plimpton, Fast Parallel Algorithms for Short-Range Molecular Dynamics, J. Comput. Phys. 117 (1995) 1–19. doi:10.1006/jcph.1995.1039.

[29]    A. Stukowski, Visualization and analysis of atomistic simulation data with OVITO–the Open Visualization Tool, Model. Simul. Mater. Sci. Eng. 18 (2010) 015012. doi:10.1088/0965-0393/18/1/015012.

[30]    P.M. Larsen, S. Schmidt, J. Schiøtz, Robust structural identification via polyhedral template matching, Model. Simul. Mater. Sci. Eng. 24 (2016) 055007. doi:10.1088/0965-0393/24/5/055007.

[31]    D. Sheppard, P. Xiao, W. Chemelewski, D.D. Johnson, G. Henkelman, A generalized solid-state nudged elastic band method, J. Chem. Phys. 136 (2012) 074103. doi:10.1063/1.3684549.

[32]    W. Xu, P.E.J. Rivera-Díaz-del-Castillo, W. Wang, K. Yang, V. Bliznuk, L.A.I. Kestens, S. van der Zwaag, Genetic design and characterization of novel ultra-high-strength stainless steels strengthened by Ni3Ti intermetallic nanoprecipitates, Acta Mater. 58 (2010) 3582–3593. doi:10.1016/j.actamat.2010.02.028.

[33]    W. Xu, P.E.J. Rivera-Díaz-del-Castillo, W. Yan, K. Yang, D. San Martín, L.A.I. Kestens, S. van der Zwaag, A new ultrahigh-strength stainless steel strengthened by various coexisting nanoprecipitates, Acta Mater. 58 (2010) 4067–4075. doi:10.1016/j.actamat.2010.03.005.

# Tunability of martensitic behavior through nanoprecipitates and other nanostructures: Supplementary Material

Samuel Temple Reeve[a], Karthik Guda Vishnu[a], Alexis Belessiotis-Richards[b], and Alejandro Strachan[a †]

[a]School of Materials Engineering and Birck Nanotechnology Center,

Purdue University, West Lafayette, Indiana 47906 USA

[b]Department of Materials,

Imperial College London, Exhibition Road, London SW7 2AZ, United Kingdom

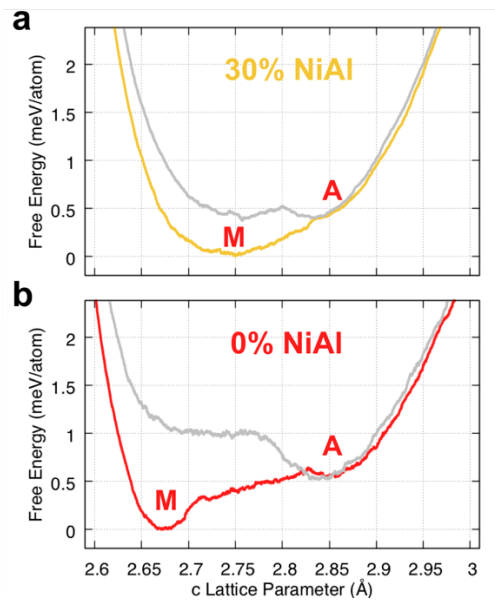

**Supplementary Figure 1:** Example free energy landscapes from nanowires starting from both the martensite (M, yellow/red), as in the main paper, and from the austenite (A, grey curves) for a) 30 at. % NiAl D-Core and b) 0 at. % NiAl nanowires. Beginning from the martensite, the system finds the austenite, but shows weak metastability. Beginning from the austenite, the system transforms to a partially martensitic structure with higher free energy.

---

[†] Corresponding Author: strachan@purdue.edu

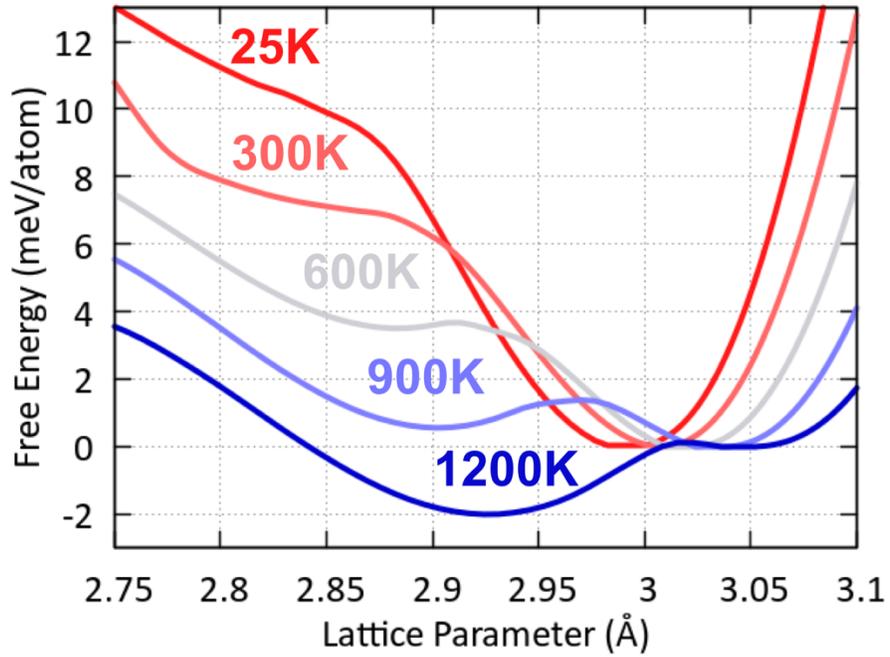

**Supplementary Figure 2:** Free energy curves for representative temperatures in the bulk (0 at. % NiAl) martensitic alloy. With increasing temperature, martensite gradually shifts from stable to metastable, while the austenite changes from unstable, to metastable, to stable.